\documentstyle[12pt]{article}

\def\xsla#1{#1 \hskip -0.19 cm / } 
\def\xxsla#1{#1 \hskip -0.22 cm / }
\def\brk{\hfill\\}

\begin{document}

\begin{flushright}
VPI-IPPAP-98-9\\
November 1998\\
\end{flushright}

\begin{center}
{\small Talk presented at ECT*/TJNAF Workshop on Excited Baryons\\
May 1998, Trento, Italy\\
To be published in \textit{Few-Body Systems Supplementum}}\\
\bigskip
\bigskip
{\Large 
Effective Lagrangian Study of $\gamma p \to K^+ \Lambda$\\
(Spin 3/2 Resonances and Their Off-Shell Effects)\\
}
\bigskip
\smallskip
\textsc{\large T. Mizutani}\\
\medskip
\textit{\normalsize 
Institute of Particle Physics and Astrophysics\\
Department of Physics, Virginia Tech, Blacksburg, VA 24061-0435, USA}\\
\end{center}

\medskip

\begin{abstract} 
The purpose of the present discussion is to
supplement the talk, by B.~Saghai at this workshop, on the study of the
electromagnetic production of strangeness on the nucleon based upon
effective Lagrangian methods.  Here we focus on the proper treatment of
the spin 3/2 resonances and their associated effects due to the spin 1/2
component of the corresponding fields when they are {\it off the mass shell}. 
\end{abstract}

\bigskip
\begin{flushleft}
\textbf{\large 1. Introduction}
\end{flushleft}

To date, in most works exploiting the effective Lagrangian approach to
investigate the photo- (and electro-) production of strangeness on the
nucleon, one includes only the tree level contributions consisting of the
$s$-channel nucleon and its excited resonant state exchanges, the
$u$-channel hyperonic resonance exchanges, and a few 
$t$-channel strange meson exchanges. 
Note that this is also known as the {\it isobar} approximation.  
As the $s$-channel incident photon energy becomes
higher, the more massive baryonic resonances with spin higher than 1/2
need to be included in this scheme.  Until recently, the inclusion of
higher spin resonances had never been exercised except in the work of
Renard and Renard \cite{Ren} in the early 70's.  However, this work
did not recognise the propagation of the spin 1/2 components of the
Rarita-Schwinger spin 3/2 field {\it off-shell}.  A recent approach
\cite{SL} took a somewhat different point of view and included spin 3/2
and 5/2 resonances, but only in the $s$-channel due to an unwanted
singularity in the $u$-channel arising from
an approximate treatment of  these higher spin objects while respecting gauge
invariance. 

A proper spin 3/2 resonance treatment was discussed \cite{RPI1} and
exploited by the Rensselaer Polytechnic (RPI) group for the photo-
(electro-) production of the pion in the $\Delta(1232)$ resonance region
\cite{RPI2}. 
In what follows we shall extend their approach to K-meson production.  
Our result in the 
\begin{displaymath} 
\gamma^{(*)} p \to K^+ \Lambda 
\end{displaymath} 
reaction has recently been published in \cite{offshell} and the
details can be found therein.  In the present discussion we shall include
one additional aspect in its yet exploratory stage, 
but not investigated in \cite{offshell}.

\bigskip
\begin{flushleft}
\textbf{\large 2. Spin 3/2 Resonances and Their Off-shell Effects}
\end{flushleft}

First, we refer the reader to appropriate articles on the proper treatment
of spin 3/2 fields for more detail \cite{RPI1,Pasc,Hemm}. 
 
Following the convention found in Bjorken and Drell \cite{BD}, we
write the spin 3/2 (isospin 1/2 for our present discussion) 
Rarita-Schwinger [vector-spinor field] (resonance) as $R^{\mu}$. 
Without interaction this field should obey the free Dirac equation
(with $M_R$: the mass of the resonance)  
\begin{equation}
(i\xxsla{\partial} - M_R)R^{\mu} = 0,
\end{equation} 
with the
subsidiary condition to be satisfied by the {\it on-shell} spin 3/2
resonance: 
\begin{equation} 
\gamma_{\mu} R^{\mu} = 0 
\end{equation}
ensuring the correct number of spin components. (The often
mentioned additional condition: $\partial_{\mu} R^{\mu} = 0$, is
automatically met by the two equations above.) Then the most general form
of the corresponding free Lagrangian reads,
\begin{equation} 
{\cal L}_{\rm free} = \overline R^{\alpha} \Lambda_{\alpha\beta}R^{\beta}, 
\end{equation}
where 
\begin{eqnarray} 
\Lambda_{\alpha \beta} &=&
-\biggl[ (-i\xxsla{\partial} + M_R)g_{\alpha \beta} -iA (\gamma_{\alpha}
\partial_{\beta} +\gamma_{\beta} \partial_{\alpha})
-\frac{i}{2}(3A^2+2A+1)\gamma_{\alpha} \xxsla{\partial} \gamma_{\beta}
 \biggr.
\nonumber \\
& & \qquad \biggl.
- M_R(3A^2+3A+1) \gamma_{\alpha} \gamma_{\beta} \biggr].
\end{eqnarray}
Here $A(\ne -1/2)$ is a free parameter. Since this Lagrangian
can be shown to be invariant under the following point (or contact)
transformation
\begin{equation} 
R^{\mu} \to R^{\mu} + a\gamma^{\mu} \gamma^{\nu} R_{\nu};\quad 
A \to A + (A-2a)/(1+4a),
\end{equation}
($a \ne -1/4$, but otherwise arbitrary), observables resulting
from this Lagrangian is free of parameter $A$.  Note that on the right
hand side of the above transformation for $R^{\mu}$, the part proportional
to $\gamma^{\nu} R_{\nu}$ may be easily seen to behave as a Dirac spin 1/2
field.  Thus {\it off the mass shell} the Rarita-Schwinger field $R^{\mu}$
is always mixed with spin 1/2 components.  This has an important bearing
on what we are going to discuss in the following. 
 
First, the propagator for the $R^{\mu}$ field is the inverse of
$\Lambda_{\mu \nu}$, and its simplest form may be found by setting $A=-1$,
\begin{equation} 
P_{\mu \nu}(q) = 
\frac{ \xsla{q} + M_R }{ 3(q^2 - M_R^2) }
\left[ 3g_{\mu \nu} -\gamma_{\mu} \gamma_{\nu} - \frac{2q_{\mu}
        q_{\nu}}{M_R^2} - \frac{q_{\nu} \gamma_{\mu} -q_{\mu}
        \gamma_{\nu}}{M_R}
\right], \label{e:prop32} 
\end{equation}
where $q$ is the four momentum of the resonance.  It is
important to note \cite{RPI1} that this propagator describing the on- and
off-shell propagation of the spin 3/2 resonance contains a spin 1/2
contribution: only at the pole of the propagator (the on-shell point) does
the spin 1/2 contribution vanish and only the pure spin 3/2 component
retained.  It has been shown that dropping the spin 1/2 component would
lead to a propagator with no inverse \cite{RPI1}. 

Next, the interactions associated with the spin 3/2 field $R^{\mu}$ which
respects the invariance under the point transformation of the free
Lagrangian must be of the form: 
${\cal L}_{I}= j^{\mu}\Theta_{\mu\nu}R^{\nu} + h.c.$, 
where the most general form for $\Theta$ was found to
be \cite{Pasc,Nath}
\begin{eqnarray} 
\Theta_{\mu \nu}(V)
& = & (g^{\alpha}_{\mu} + V\gamma_{\mu}\gamma^{\alpha})
(g^{\alpha}_{\nu}+A/2\cdot\gamma^{\alpha}\gamma_{\nu}) \cr
& \to & g_{\mu \nu} -
(V+1/2)\gamma_{\mu}\gamma_{\nu}\qquad (A= -1).\phantom{\frac{1}{2}}
\label{eq:teta} 
\end{eqnarray}
Here $V$ is a free parameter, and while the physical quantities
are independent of $A$, as the full Lagrangian is invariant
under the point transformation \cite{KOS}, they do depend upon $V$. From
the above expression one sees that $V$ enters to modify only the spin 1/2
part of the Rarita-Schwinger field.  Hence, it is clear that the spin 3/2
resonance pole part of the amplitudes is independent of $V$. For
this reason $V$ is called the {\it Off-Shell Parameter}.  
Our objective then is 
to study the effect of these off-shell degrees of freedom 
on the electromagnetic production of strangeness on the nucleon.  
Note that this cannot be done consistently while
using an inappropriate resonance propagator which
ignores the spin 1/2 contribution.  
Since our spin 3/2 fields are not
elementary but composites of more fundamental constituents, the
corresponding off-shell parameters should not be fixed to definite
values: a point of view adopted by \cite{RPI1}. 

Our phenomenological interaction Lagrangian consists of the following
pieces: 
\begin{eqnarray} 
{\cal L}_{K\Lambda R} 
& = & \frac{g_{K\Lambda R}}{M_K}
\left[ \overline{R}^{\nu} \Theta_{\nu\mu}(Z) \Lambda \partial^{\mu} K +
       \overline{\Lambda}(\partial^{\mu} K^{\dagger})\Theta_{\mu\nu}(Z)R^{\nu}
\right],  \label{eq:intlag1} \\  
{\cal L}_{\gamma p R}^{(1)} 
& = & \frac{i e g_1}{2 M_p} 
\left[ \overline{R}^{\nu} \Theta_{\mu\lambda}(Y) \gamma_{\nu}\gamma^5 N
       F^{\nu\lambda}
     + \overline{N}\gamma^5\gamma_{\nu} \Theta_{\lambda\mu}(Y)
       R^{\mu} F^{\nu\lambda} 
\right], \label{eq:intlag2} \\  
{\cal L}_{\gamma p R}^{(2)} 
& = & \frac{- e g_2}{4 M_p^2} 
\left[ \overline{R}^{\mu} \Theta_{\mu\nu}(X) \gamma^5 (\partial_{\lambda} N) 
       F^{\nu\lambda}
     - (\partial_{\lambda} \overline{N}) \gamma^5 \Theta_{\nu\mu}(X) R^{\mu}
       F^{\nu\lambda} 
\right],  \label{eq:intlag3} \\ 
{\cal L}_{\gamma p R}^{(3)} 
& = & \frac{- e g_3}{4 M_p^2} 
\left[ \overline{R}^{\mu} \Theta_{\mu\nu}(W) \gamma^5N (\partial_{\lambda}
       F^{\nu\lambda})
     - (\partial_{\lambda} F^{\nu\lambda})\overline N \gamma^5
       \Theta_{\nu\mu}(W)  R^{\mu} 
\right]. \cr
& & \label{eq:intlag4} 
\end{eqnarray}
The above expression may be found in \cite{offshell} except for
the last piece, the necessity of which has been communicated to us by Rick
Davidson \cite{David}.  This part contributes only to the
electroproduction process, which may be obvious from the presence of the
divergence of the electromagnetic tensor $F^{\nu\lambda}$ in the
interaction: classically this quantity vanishes when there is no source
for electric charge/current. 

\bigskip
\begin{flushleft}
\textbf{\large 3. Result and Discussion}
\end{flushleft}

With the introduction of the spin 3/2 resonances as discussed above, 
we constructed the tree level amplitude for 
$\gamma^{(*)} p \to K^+ \Lambda$,
and performed a $\chi^2$ fit to the existing data in which the coupling
strengths of the intermediate resonances (to the initial and final states,
in combination) and the off-shell parameters: $W, X, Y, Z$ were varied.
  
The result without the ${\cal L}_{\gamma p R}^{(3)}$ term has been worked
out in \cite{offshell}, and the reader is referred to that article for
details.  
In summary, the correct treatment of the spin 3/2 resonances has
improved the fit to the existing data. 
In particular, the undesirable
increase in the integrated cross section for $E^{Lab}_{\gamma} > 1.5$ GeV
in previous models (\cite{SL}, for example) has been considerably
tempered, and the large angle differential cross section at
$E^{Lab}_{\gamma} = 2.0$ GeV has changed drastically, in favor of the yet
preliminary data from Bonn \cite{BIGSKY}.  
Predictions for several
polarisation observables both in photo- and electro-productions show a
marked difference between the approaches with and without the off-shell
considerations in spin 3/2 resonances.  

Regarding the off-shell effects
arising from the ${\cal L}_{\gamma p R}^{(3)}$ term which only affects
electroproduction, our preliminary results indicate a rather minor effect:
we basically need ample data points to firmly constrain the corresponding
parameter(s): coupling strength and the off-shell parameter $W$. 

Incorporating resonances with spin higher than 3/2 is yet to be
investigated within the context of covariant schemes.

\bigskip
\begin{flushleft}
\textbf{\large Acknowledgements}
\end{flushleft}
The content of this note is a result of a pleasant
collaboration with Claude~Fayard, Georges~Lamot, and Bijan~Saghai.  
We would like to thank Rick~Davidson for pointing out
the possibility of the third electromagnetic coupling term 
${\cal L}_{\gamma p R}^{(3)}$.
The warm hospitality of the ECT$^*$ at Trento during the workshop 
is gratefully acknowledged by the present author.


\end{document}